# *The Problem of the Age of the Universe and the Earliest Galaxies*

*Roger Ellman*

Abstract

A number of years ago the estimates of astronomers and astrophysicists were that the earliest galaxies took about 2½ - 3 billion years to form, that is, that they did not appear until *2.5–3.0 billion years* after the Big Bang. Those estimates were based on analysis of the processes involved in star formation and in the aggregation and "clumping" of matter in the early universe.

Since then improved equipment and techniques [e.g. Keck and Hubble telescopes and gravitational lensing] have resulted in reports of observation of early galaxies having stars that formed as early as *300 million years* after the Big Bang.

Such new data has led to the abandonment of the several billion years estimates of the time required for star and galaxy formation; however, an alternative response to the data would be to re-examine the Hubble theory from which the age of the universe and the distance to high redshift objects is determined.

Roger Ellman, The-Origin Foundation, Inc.
320 Gemma Circle, Santa Rosa, CA 95404, USA
RogerEllman@The-Origin.org
http://www.The-Origin.org

# *The Problem of the Age of the Universe and the Earliest Galaxies*

*Roger Ellman*

## *Background of the Problem*

A number of years ago the estimates of astronomers and astrophysicists were that the earliest galaxies took about 2½ - 3 billion years to form, that is, that they did not appear until *2.5-3.0 billion years* after the Big Bang. Those estimates were based on analysis of the processes involved in star formation and in the aggregation and "clumping" of matter in the early universe.

Since then improved equipment and techniques [e.g. Keck and Hubble telescopes and gravitational lensing] have resulted in reports of observation of early galaxies having stars that formed as early as *300 million years* after the Big Bang.

Such new data has led to the abandonment of the several billion years estimates of the time required for star and galaxy formation; however, an alternative response to the data would be to re-examine the Hubble theory from which the age of the universe and the distance to high redshift objects is determined.

Analysis of the Hubble Law shows that it is asymptotic to an age of the universe that depends on the value of the Hubble Constant. The value of the Hubble Constant is generally taken as in the range of *60 to 75*, but its value remains to be determined. The current generally accepted age of the universe is *13.7 billion years*, which corresponds to a Hubble Constant of *67*. The most recent [2012] determination of a value for the Hubble Constant is $74.3 \pm 2.1$.

For high $z$ cosmic objects the Hubble Law results in recession velocities approaching the speed of light. That those velocities are attributed to expansion of space, not to actual velocity of the objects, does not really relieve the problem. According to the Hubble Law the distance between we the observers and those far distant cosmic objects is nevertheless increasing at a rate almost the speed of light which is unreasonable for such immense masses.

The problem of sufficient time after the Big Bang for stars to form, the unreasonable recession velocities implied by the Hubble Law, and even that the Hubble "constant", on which those all depend, is so poorly determined and appears to not be subject to better determination, would all evidence that the Hubble theory is defective and should be replaced.

## *The Asymptotic Hubble Law*

### The Effect of the Hubble Constant on the Implied Age of the Universe

*(1)* Where: $c \equiv$ light speed.
$v \equiv$ observed astral body's recession velocity.
$d \equiv$ distance of observed astral body away from us.
$z \equiv$ observed redshift.
$H \equiv$ Hubble Constant.

And: $v = H \cdot d$ or $H = {}^{v}/_{d}$ ["Hubble Law"]

*(2)* $z \equiv$ relativistic redshift due to the Doppler effect

$$= \frac{[1 + {}^{v}/_{c}]^{½}}{[1 - {}^{v}/_{c}]^{½}} - 1$$

$$(3) \qquad v = c \cdot \frac{[z+1]^2 - 1}{[z+1]^2 + 1} \qquad \text{[solve (2) for “v”]}$$

$$(4) \qquad d = \frac{v}{H} \qquad \text{[solve Hubble Law of (1) for "d"]}$$

$$= \frac{c}{H} \cdot \frac{[z+1]^2 - 1}{[z+1]^2 + 1} \qquad \text{[substitute "v" from (3) above]}$$

This $d$, the distance of the observed light source, and the time ago that its light that we now observe was first emitted, is asymptotic to $c/H$. Therefore, the most distant, the oldest light source possible is $c/H$. Calculating that in billions of light years for various values of $H$ the following values for the “age of the universe” are obtained.

(5) Using: $c = 3 \cdot 10^8$

For H = 63 “age of the universe” = 14.6 billion years

For H = 75 “age of the universe” = 12.3 billion years

For H = 88 “age of the universe” = 10.5 billion years

or

For “age of the universe” = 13.7 billion years H = 67.

This asymptotic behavior of the Hubble Law is illustrated in Figure 1, below.

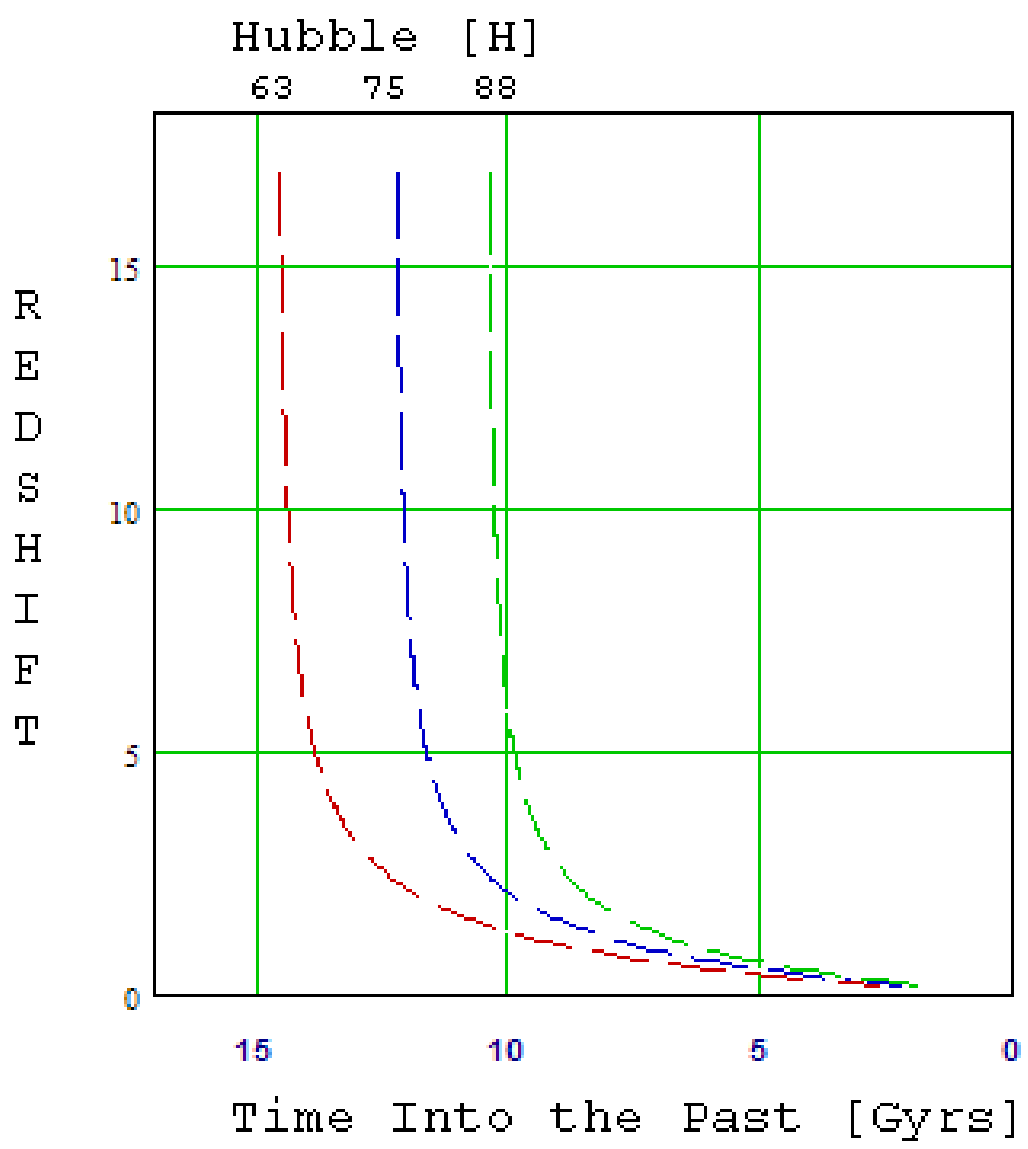

*Figure 1 - The Asymptotic Hubble Law*

The Hubble Law embodies an interpretation of redshifts and a system of distance measurement to cosmic objects. An alternative to it that nevertheless fulfills those functions develops as follows.

## *An Alternative to the Hubble Law*

Given two particles [e.g. electrons or protons] that have electric charge, the particles being separated and with the usual electric [Coulomb] force between them, if one of the charged particles is moved the change can produce no effect on the other charge until a time equal to the distance between them divided by the speed of light, $c$, has elapsed.

For that time delay to happen there must be something flowing from one charge to the other at speed $c$ and each charge must be the source of such a flow.

That electric effect is radially outward from each charge, therefore every charge must be propagating such a flow radially outward in all directions from itself, which flow must be the "electric field".

Likewise, given two masses, i.e. particles [e.g. electrons or protons] that have mass, being separated and with the usual gravitational force [attraction] between them, if one of the masses is moved the change can produce no effect on the other mass until a time equal to the distance between them divided by the speed of light, $c$, has elapsed.

For that time delay to happen there must be something flowing from one mass to the other at speed $c$ and each particle, each mass must be the source of such a flow.

That gravitational effect is radially outward from each mass, therefore every mass must be propagating such a flow radially outward in all directions from itself, which flow must be the "gravitational field".

Therefore, the fundamental particles of atoms, of matter, which have both electric charge and gravitational mass, must have something flowing outward continuously from them and:

· Either the particles have two simultaneous, separate outward flows, one for the effects of electric charge and another for gravitation, or

· They have one common universal outward flow that acts to produce both of the effects: electric field and gravitational field.

There is clearly no contest between the alternatives. It would be absurd for there to be two separate, but simultaneous, independent outward flows, for the two different purposes. They both are flow of the same medium – a universal outward flow.

But, for there to be a continuous flow outward from each particle, each must be a supply, a reservoir, of that medium which is flowing. The original supply of the flow medium came into existence at the "Big Bang" the beginning of the universe.

Since the original "Big Bang" the outward flow has been inevitably gradually depleting the original supply. That process, an original quantity gradually depleted by flow away of some of the original quantity is an exponential decay process and the rate of the decay is governed by its time constant.

Development and elaboration of the details of that overall universal decay is fully developed in reference[1]. In summary, the Universal Exponential Decay is a decay of the length dimensional aspect of all quantities in the universe. It involves the fundamental constants ($c$, $q$, $G$, $h$, etc.) and decay of any of those must be, and is, dimensionally consistent with the decay of the others.

The dimension that is decaying is length, the $[L]$ dimension in the dimensions of, for example: the Planck Constant, $h$, $[M \cdot \mathbf{L}^2/T]$; the speed of light, $c$, $[\mathbf{L}/T]$; and the Newtonian Gravitational Constant, $G$, $[\mathbf{L}^3/M \cdot T^2]$. The time constant of the decay, $\tau$, is as shown in equation (6).

$$(6) \qquad \tau = 3.57532 \cdot 10^{17} \text{ sec } (\approx 11.3373 \cdot 10^9 \text{ years})$$

The actuality of the Universal Decay is validated by a number of otherwise unexplained anomalies including galactic rotation curves, the Pioneer anomaly, and the flybys anomaly[2].

## *The Universal Decay and RedShift*

The initial outward flow, carrying electric and gravitational field outward from the original product particles of the "Big Bang", flowed into empty space, space unoccupied by anything that could set its speed of propagation. Therefore, the speed of propagation of field from those initial particles, and such particles in general, must be set by the speed at which propagated by / from the particle. The speed determining factors have to be part of the flow, of what is flowing, since there is no where else they could come from.

Therefore, while the Universal Decay was steadily gradually reducing the speed of propagation of the outward flow, earlier increments of propagated flow continued at their original speed, as set by the speed at which they were initiated and as maintained by the characteristics of the increment of flow included in them when then propagated.

Comparison of light propagated early in the Universal Decay against local "Earth" light discloses that the earlier light is of longer wavelength, longer because less decayed. That is the major factor in redshift. There is undoubtedly some redshift due to the doppler effect, but the principal component of redshifts is that due to how much less decayed the observed light is compared to local contemporary light. See the third reference.

Then, how old is the oldest, least decayed propagation; how old is the universe ?

## *The Age of the Universe*

Figure 2, below is a plot of the Universal Exponential Decay versus redshift per the decay time constant of equation (6). [The figure also illustrates how negligible is the doppler redshift component of the total redshift, analyzed and determined in reference[3].]

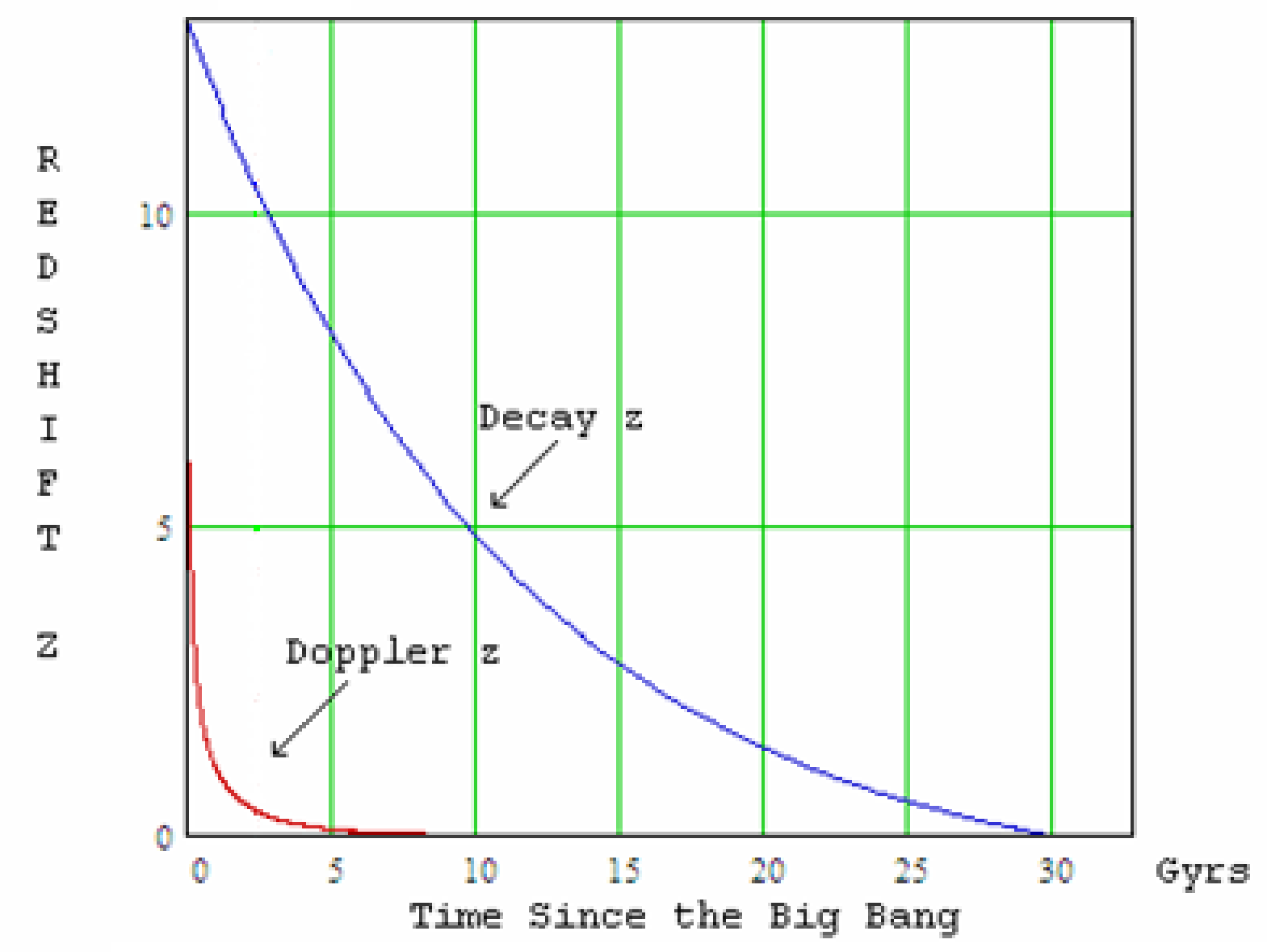

*Figure 2*
*Redshifts: Caused by the Universal Decay [and by the Doppler Effect]*

Redshifts as high as $z = 10$ have been reported and there have been indications of redshifts of $z = 12$. Further improvements in equipment and techniques may result in observation of even higher redshifts.

Figure 2 makes clear why the age of the universe must be on the order of $30\ Gyrs$ or more. That amount of time is needed to include enough time constant periods, $\tau$, that is $11.3373\ Gyrs$ each, to yield, to make possible, an observed redshift of $z > 10$. The value for the present estimated age of the universe, $age = 30\ Gyrs$ is an, at present, conservative best estimate taking into account current observational data.

This Universal Exponential Decay completely resolves the problem of sufficient time after the Big Bang for the earliest stars to form. No matter how high the observed $z$ may be the beginning of the exponential decay preceded it by the required star formation time. We may never know how long ago the Big Bang happened and we may never know the requisite time for the earliest stars to form, but we can know how long ago the oldest stars observed formed and we can know that the Big Bang took place before then.

Figure 3 graphically compares the Universal Exponential Decay and the Hubble Theory to the same scale, each having been independently presented in Figures 1 and 2.

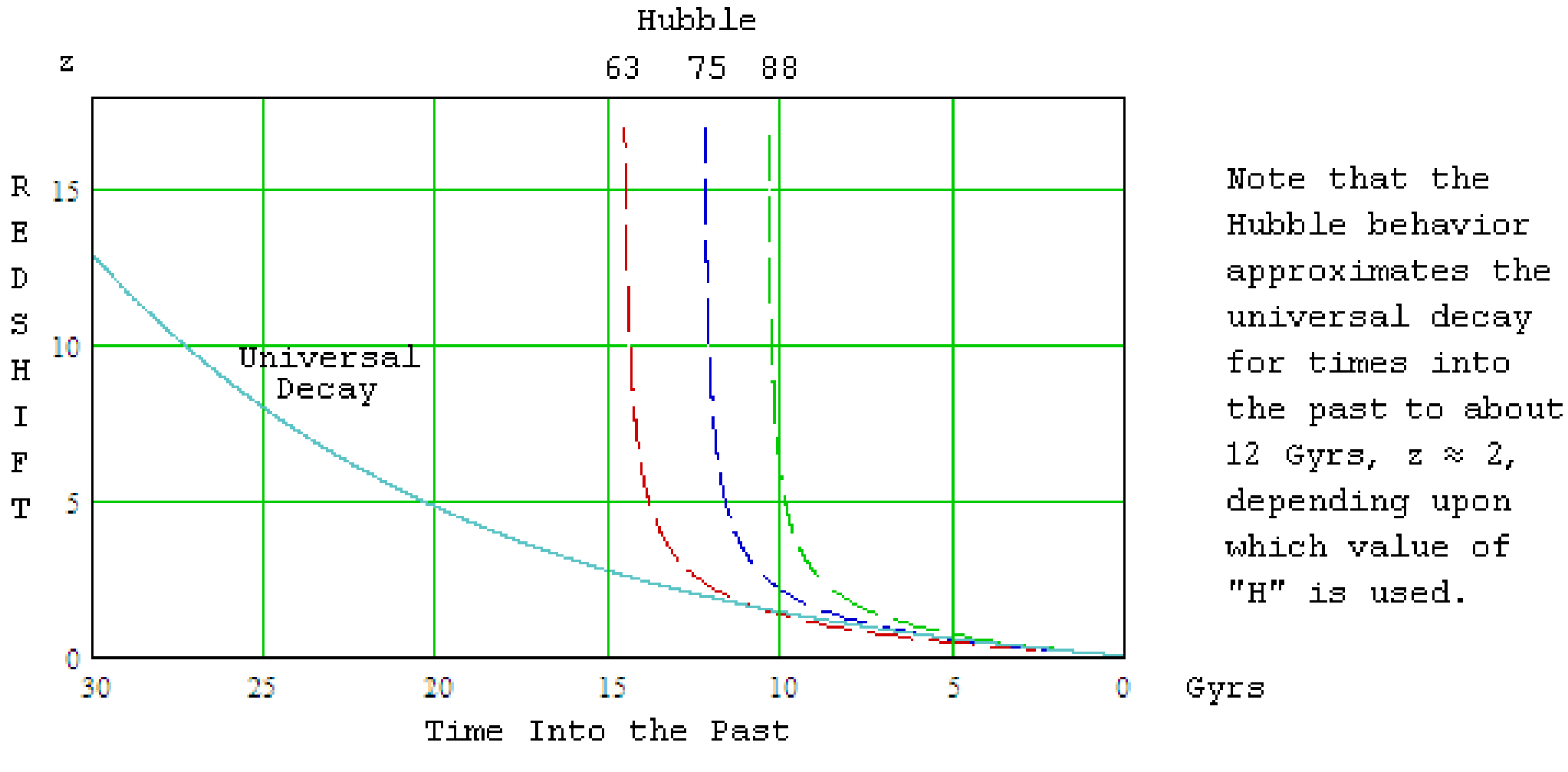

*Figure 3*

*Comparison of the Universal Exponential Decay vs. the Hubble Theory*

## Conclusion

The Hubble theory, while useful during its early years, has become a problem and a handicap and is interfering with the progress of cosmology, astrophysics and astronomy. As with another theory that was useful in its early days even though profoundly in error, namely Ptolemy's solar-centric universe system, the Hubble theory should be dropped and its place taken by the Universal Exponential Decay.

The Universal Exponential Decay not only works better than the Hubble theory, it is also much more well validated by other phenomena than is the Hubble theory[2].